# How water meets a very hydrophobic surface

Sudeshna Chattopadhyay,[1] Ahmet Uysal,[1] Benjamin Stripe,[1] Young-geun Ha,[2] Tobin J. Marks,[2] Evguenia A. Karapetrova,[3] and Pulak Dutta[1*]

[1]Dept. of Physics & Astronomy, Northwestern University, Evanston, IL 60208, USA
[2]Dept. of Chemistry, Northwestern University, Evanston, IL 60208, USA
[3]Advanced Photon Source, Argonne National Laboratory, Argonne, IL 60439, USA

Is there a low-density region ('gap') between water and a hydrophobic surface? Previous X-ray/neutron reflectivity results have been inconsistent because the effect (if any) is sub-resolution for the surfaces studied. We have used X-ray reflectivity to probe the interface between water and more hydrophobic smooth surfaces. The depleted region width increases with contact angle and becomes larger than the resolution, allowing definitive measurements. Large fluctuations are predicted at this interface; however, we find that their contribution to the interface roughness is too small to measure.

The hypothesis that density-depleted regions exist between water and hydrophobic surfaces has been actively discussed for several decades.[1-6] It is a compelling idea for many reasons, for example it provides a simple explanation for the large slip lengths seen in studies of shear flow at hydrophobic surfaces.[7] A detailed understanding of hydrophobicity is essential to understanding how proteins fold in aqueous environments, membrane-water interactions, and the physics of interfaces within colloids. Therefore there have been numerous efforts to study the water-hydrophobic interface density profile with computer simulations and experiments. However, consensus has been surprisingly elusive and the picture has remained murky.

Computer simulations (reviewed in, e.g. Ref. 5-6) have produced a wide variety of very detailed results but no consistent picture. This is due in part to the use of different simulated substrates and different models, but also to the relative absence of experimental information to constrain the simulations. Neutron and X-ray reflectometry are the best available nanoscale probes of density profiles normal to a surface, but this problem pushes these techniques to their limits. The reflectivity from an interface, $R(q)$, is determined by the Fourier transform of $d\rho(z)/dz$, where $\rho(z)$ is the laterally averaged electron density (for X-rays) or scattering length density (for neutrons).[8] The measured intensity drops rapidly with momentum transfer $q$ and can be measured only up to a limited $q_{max}$: for typical studies of water-hydrophobic interfaces, $q_{max} \sim 0.9\text{Å}^{-1}$ for X-rays and $<0.3\text{Å}^{-1}$ for neutrons. The consequence of inverting limited-range $R(q)$ data is that features in ρ(z) are convoluted with a resolution function[9] of width $\Delta z \sim \pi/q_{max}$. (~3Å for typical X-ray studies, and >10Å for typical neutron studies). Thus narrow intrinsic features in ρ(z) are broadened, and weak features may escape detection. The substrate roughness also smears the intrinsic gap density profile.

---

[*] e-mail: pdutta@northwestern.edu

The experimental situation must be reassessed in light of the resolution limit. Several neutron reflectivity studies[10-12] have reported large depletion widths (comparable to the >10Å resolution limit). Another neutron experiment saw no depletion.[13] An X-ray reflectivity study of a liquid-liquid interface,[14] with $q_{max}$ restricted to 0.4Å$^{-1}$ (resolution limit ~7A) also reported a null result. The two highest-resolution X-ray studies to date are by Poynor et al[15] using octadecyltriethoxy-silane (OTE) self-assembled monolayers (SAMs), and Mezger et al.[16] using octadecyltrichloro-silane (OTS) SAMs. Both SAMs are identically methyl-terminated at the upper surface. The observed depletion layer widths were 2-4Å (Ref. 13) and 1-6Å (Ref. 14). In these cases the resolution limit was ~3Å. Thus while the X-ray widths are smaller than the neutron widths, the measurements are still resolution-limited.

Recognizing the problem, Mezger et al.[16] determined the product of the depleted zone width $D$ and the electron density *deficit* relative to bulk water $\rho_{H2O}-\rho_{dep}$, and found this product (the integrated depletion) to be a more robust parameter. Dividing by $\rho_{H2O}$ yields a more intuitive variable:[17] the equivalent width of a zero-density gap $D_{eq} \equiv D(\rho_{H2O}-\rho_{dep})/\rho_{H2O}$. The data in ref. 15 and 16 yield $D_{eq}$=1.8-3.3Å and $D_{eq} \approx 1.0$Å (no error bar reported) respectively. It may appear that both these results qualitatively confirm the presence of a depleted region even if they differ quantitatively. However, Ocko et al[18] have pointed out that on a dry SAM, the terminal hydrogen atoms (which have few electrons) appear to X-rays as part of the ambient air or vacuum, but on a wet SAM the hydrogen layer appears as an intervening electron-depleted region with $D_{eq} \approx 1.0$Å. Indeed Poynor et al[19] find that the ethanol-OTE interface also shows a depletion (using their numbers, we find $D_{eq} \approx 1.0$Å), even though the contact angle of ethanol on OTE is small. Once the spurious "hydrogen gap" is subtracted, the data in ref. 13 show no gap, while the data in ref. 15 show a gap with corrected $D_{eq}$~0.8-2.3Å. A detailed neutron reflectivity study by Maccarini et al.[17] also observed depletion both with water and with nonpolar liquids, and the difference has error bars that include 0. For example, at pH=5.5, $D_{eq}$=1.4 ± 1.6Å.

We suggest that the qualitative and quantitative differences between various reports do not indicate a substantive controversy; they are merely scatter in measurements of sub-resolution numbers. If so, how can these crucial measurements be made definitive? In these systems a mere two-fold improvement in resolution would require several orders of magnitude increase in usable intensity, which is impractical. The other approach is to seek a stronger and more easily measurable effect.

We have studied the interface between water and fluoroalkylsilane SAMs. Specifically, we have fabricated $CF_3(CF_2)_5(CH_2)_2SiCl_3$ and $CF_3(CF_2)_{11}(CH_2)_2SiCl_3$ SAMs, referred to below as FAS13 and FAS25 respectively. Fluorocarbon-based SAMs are not necessarily very hydrophobic; the contact angle depends on chain length.[20] The advancing contact angle for our samples was only 111° for FAS13, but 120° for FAS25. This trend is the same as that previously reported.[20] A close-packed array of $CF_3$ groups is thought to be the most hydrophobic smooth surface achieva-ble in nature,[21] and the observed contact angle for epitaxially grown $CF_3$-terminated surfaces[21] is ~120°, which is close to our FAS25 result. Rough or microstructured (superhydrophobic) surfaces can have larger contact angles, but the observed surface roughness of our SAMs is only ~3Å RMS (i.e. resolution-limited). Larger contact angles can also be generated in computer simulations of smooth surfaces,[22,23] but it has been shown that these do not reproduce measured contact angles on the physical surfaces that they aim to simulate.[23]

[2]

The terminal atoms at these surfaces are fluorine (Z = 9) rather than hydrogen. We estimate that a slab containing only the top layer of F atoms has roughly half the electron density of water. We do not use such a separate slab in our fits, but the terminal atoms contribute to the profile of the SAM surface, and are not "invisible".

Each wet sample in the present study (Fig. 1, upper panel inset) consists of a film of water between the SAM-coated silicon substrate and a 7.5μm thick Kapton window.[12,24] Water has a stronger tendency to bead on more hydrophobic surfaces, but a sufficiently hydrophilic window will make a uniform sandwiched water film energetically favorable. After the Kapton was made more hydrophilic (and also cleaned) by treatment in an oxygen plasma for 3 minutes, ellipsometry confirmed that ~1μm thick uniform water films are formed.

FAS13 was purchased from Aldrich and FAS25 from Synquest Inc; both were used as supplied. SAMs were deposited using the methods of Paso et al.[20], except we also repeatedly filtered the solution because undissolved material causes rougher surfaces. X-ray reflectivity studies were performed at Sector 33-BM-C of the Advanced Photon Source. The X-ray energy was 19KeV, and the beam size was ~0.4mm vertically and ~1mm horizontally. The vertical momentum resolution was ~0.007Å$^{-1}$. No area of the sample was exposed to X-rays for more than ~1 hour, and over this time period, successive reflectivity scans were identical. Radiation damage effects become apparent after longer exposure times.

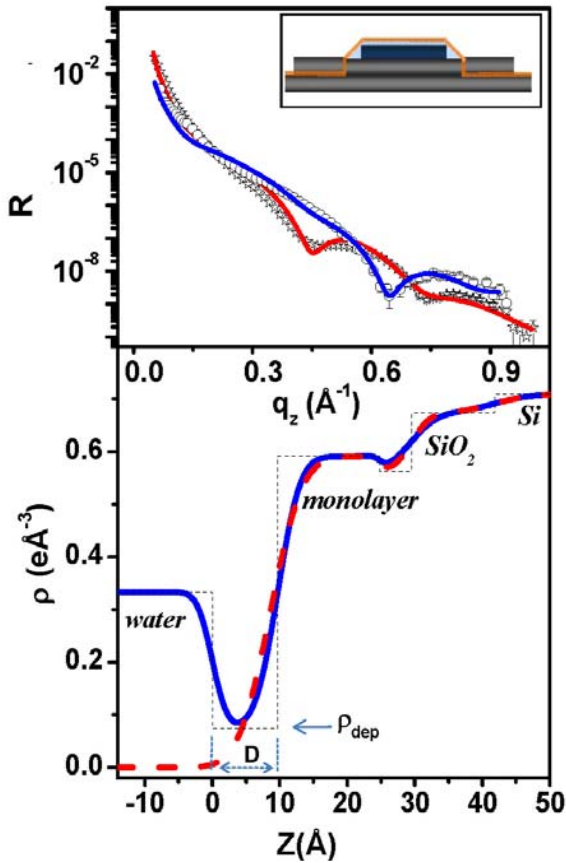

Fig.1. *Upper panel:* Unnormalized reflectivity data, and best fits (lines through data), for the dry FAS25 SAM surface (stars) and for the SAM-water interface (circles). We show R rather than $R/R_F$ for consistency with other papers in this area.[15-17,23] *Inset*: schematic diagram of the sample cell, shown in cross-section. X-rays penetrate a Kapton window and are reflected from the water-substrate interface. *Lower panel*: fitted density profile of the dry FAS25 SAM (dashed line) and wet FAS25 SAM (solid line). Slab thickness variations much smaller than the resolution width (~3Å) are not significant. The thin dashed lines show the unrounded slabs for the wet SAM.

[3]

The reflectivity data were fitted using the Parrat formalism,[25] where the electron density profile is modeled as a series of uniform-density slabs connected by rounded interfaces. Known parameters such as water and silicon densities were fixed. For the dry SAM, the model contained slabs for the semi-infinite silicon substrate, the oxide layer, and two regions of the SAM. The region closest to the substrate accounts for the silane region[15,16] and the $(CH_2)_2$ section combined. We have attempted to assign a separate slab to the subresolution $(CH_2)_2$ section, but the width goes to zero if allowed to vary. In studies of similar SAMs,[26] good fits were obtained either with or without a separate hydrocarbon slab. The insertion of slabs much thinner than the resolution is not reasonable, and results in the undesirable proliferation of insignificant fitting parameters.

The SAM-water interface reflectivity data were fitted by adding slabs for the interface region and the semi-infinite water region. Only gap parameters are reported below; the SAM parameters are provided in a supplementary document.[27] Note that the present experiments cannot see the high level of detail that is generated in some simulations. Given the limited resolution, the gap can reasonably be represented by a single slab at least as wide as the resolution. This is analogous to pixellation in photographs, and does not indicate that the gap really is a uniform-density region. Note also that the atomic number density (theoretical), scattering length density (neutrons), and electron density (X-rays) are different in principle, but these differences are obscured in practice by resolution smearing.

In Fig. 1, the top panel shows X-ray reflectivity data for the FAS25 surface and the FAS25-water interface. It can be seen, by comparison to the data in Ref. 15 and 16, that the lateral shift between the dry and wet reflectivity oscillations is much larger in our data. This indicates, independently of fitting procedures, that the effect of interest is larger in our system. The dry and wet SAM density profiles obtained from data fitting are shown in the lower panel. The dotted straight lines indicate the slabs (without interface rounding) for the wet case. We find $D = 9.7\pm0.5$Å and $\rho_{dep}=0.08\pm0.02$ electrons/Å$^3$. This gives $D_{eq} \approx 7\pm1$Å. Fig. 2 shows reflectivity data and the fitted profile for FAS13. In this case we find $D=9\pm0.5$Å and $\rho_{dep}=0.16\pm0.02$ e/Å$^3$. Thus $D_{eq}\approx4.5\pm1$Å. Recall that the contact angle for FAS13 is only ~111°.

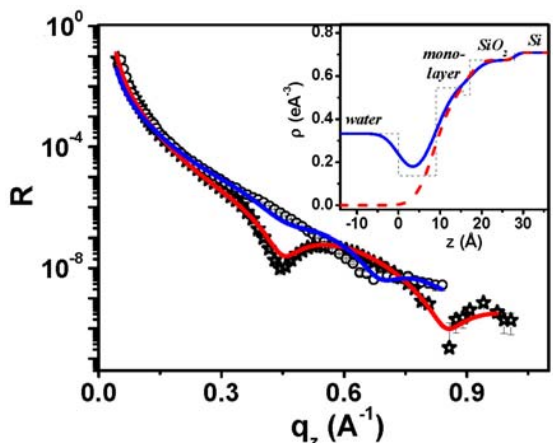

Fig. 2: Unnormalized reflectivity data, and best fits (lines through data), for dry SAM surface (stars) and SAM-water interface (circles). *Inset:* Density profile of dry FAS13 SAM (dashed line) and FAS13-water interface (solid line). Thin dashed lines show unrounded slabs for the wet SAM.

The fact that SAMs of different chain lengths, handled in the same way, give different results, suggests that the gaps are not due to extraneous factors. Our studies were conducted both with ultrapure water (from a Milli-Q system, resistivity 18.2 MΩ-cm) and less-pure deionized filtered water (resistivity 0.5 MΩ-cm), with identical results.

A facile explanation for interfacial gaps is that they may be due to dissolved gases. The gap density is low, and gases have low densities. However, the gap widths are comparable to or less than the average intermolecular distance in an ideal gas at STP, and so any analogy to three-dimensional gases is flawed. On the other hand, there is always a gas (water vapor) that degassing cannot remove, and so the hypothesis is not easily falsifiable. Our results cannot be attributed to nanobubbles, for reasons discussed in detail elsewhere.[15,16] A neutron experiment[12] reported that dissolved gases change the gap width, but higher-resolution x-ray experiments[15,16] have seen no significant effect. It has recently been reported that the water/OTS interfacial region contains adsorbed $OH^-$ and $H_3O^+$ ions,[28] and this composition change could result in intermediate densities similar to those observed.

A recent paper by Mezger et al,[23] published while our paper was under review, adds one new data point: the gap between water and FAS17 (a $CF_3(CF_2)_5(CH_2)_2SiCl_3$ based SAM). Such SAMs are not very hydrophobic: the contact angle for the actual sample studied was not reported, but other studies[20,29] find that it is only 109-112°. As expected, the observed depletion is small: D=3.2Å and $D_{eq}$=1.4Å (no error bar reported).

Fig. 3 summarizes our results along with other recent results.[15-17,23] We compare $D_{eq}$ because the gap width and density are not individually reliable for some results quoted. At the low end, the variations may be attributed to the fact that these are measurements of sub-resolution effects. For more hydrophobic surfaces, the gap is larger and can be more reliably characterized. The same trend is seen in recent simulations[30,31] but their gaps are smaller. We do not suggest that the gap width depends *purely* on the contact angle; there are undoubtedly unidentified and/or uncontrolled variables. One such variable is radiation damage, which reduces the observed gap. Our own studies were conducted using a bending magnet beam, which causes less x-ray exposure per unit area than typical undulator beams.

Finally, we turn to the profile of the water-gap interface. It has been proposed[32,33] that this interface, delicately suspended between attractive and repulsive interactions, is subject to large fluctuations driven by soft modes. Our technique gives time-averaged information and does not observe the dynamics directly, but it is sensitive to the resulting smearing of the interface. Looking at the interface between bulk water and the depleted regions, the RMS interface roughness according to our fits is 2-3Å. These numbers are comparable to the resolution, and also to the roughness of the substrate surface with which it may be partly conformal. Thus the true fluctuation amplitude is too small to be determined. The interface roughness at the free surface of water as would be observed with our q-resolution is ~3.5Å.[34,35] In other words the water density fluctuations normal to the interface are not larger, and could be smaller, than at the free surface.

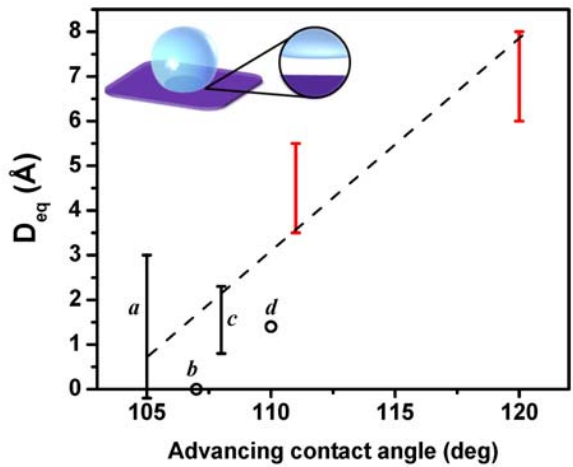

Fig. 3: Dependence of the $D_{eq}$ on advancing contact angle. *a*: from ref. 17, for partially-deuterated alkanethiol SAMs, pH = 5.5, after subtracting the gap with nonpolar liquid; *b*: from ref. 16, for OTS, after subtracting the "hydrogen gap" (error bar not reported; contact angle of sample studied not reported, value shown is estimated); *c:* from ref. 15, for OTE, also with the hydrogen gap subtracted; *d*: from ref. 23, for FAS17 (error bar not reported; contact angle of sample studied not reported, value shown is average of three contact angles reported in ref. 20 and ref. 29). The unlabeled data are from the present paper, for FAS13 and FAS25. The dashed line is a guide to the eye.

In summary, experimental limitations obscure weak or narrow intrinsic features in interfacial density profiles. We have studied a system in which the effect is stronger, and have thus obtained clear evidence of increased depletion at more hydrophobic surfaces.

ACKNOWLEDGMENTS. We thank A. Facchetti and I. Kuljanshvili for their advice and assistance, and Ali Dhinojwala, Paul Fenter, Steve Granick and Ben Ocko for detailed comments on an early draft of this paper. This work was supported by the U.S. National Science Foundation under Grant no. DMR-0705137. The Advanced Photon Source (APS) and Sector 33-BM are supported by the U.S. Department of Energy.